This manuscript has been authored by UT-Battelle, LLC, under Contract No. DE-AC0500OR22725 with the U.S. Department of Energy. The United States Government retains and the publisher, by accepting the article for publication, acknowledges that the United States Government retains a non-exclusive, paid-up, irrevocable, world-wide license to publish or reproduce the published form of this manuscript, or allow others to do so, for the United States Government purposes. The Department of Energy will provide public access to these results of federally sponsored research in accordance with the DOE Public Access Plan (http://energy.gov/downloads/doe-public-access-plan).

# Autonomous convergence of STM control parameters using Bayesian Optimization

Ganesh Narasimha[1], Saban Hus[1], Arpan Biswas[1], Rama Vasudevan[1]*, Maxim Ziatdinov[2]*

[1] *Center for Nanophase Material Sciences (CNMS), Oak Ridge National Laboratory (ORNL), Oak Ridge, Tennessee, USA – 37831*

[2] *Computational Sciences and Engineering Division (CSED), Oak Ridge National Laboratory (ORNL), Oak Ridge, Tennessee, USA – 37831*

**Abstract:**

Scanning Tunneling microscopy (STM) is a widely used tool for atomic imaging of novel materials and its surface energetics. However, the optimization of the imaging conditions is a tedious process due to the extremely sensitive tip-surface interaction, and thus limits the throughput efficiency. Here we deploy a machine learning (ML) based framework to achieve optimal-atomically resolved imaging conditions in real time. The experimental workflow leverages Bayesian optimization (BO) method to rapidly improve the image quality, defined by the peak intensity in the Fourier space. The outcome of the BO prediction is incorporated into the microscope controls, i.e., the current setpoint and the tip bias, to dynamically improve the STM scan conditions. We present strategies to either selectively explore or exploit across the parameter space. As a result, suitable policies are developed for autonomous convergence of the control-parameters. The ML-based framework serves as a general workflow methodology across a wide range of materials.

Scanning tunnelling microscopy (STM) is an invaluable tool for imaging material surfaces at atomic resolution and obtaining spectroscopic information related to their local electronic structure.[1,2] STM utilizes an atomically sharp tip, which when biased, measures currents from a conductive surface, based on the quantum mechanical tunnelling effect.[3] This technique has been used to explore various physical processes such as band-gap energetics,[4] superconductivity,[5] and other quantum phenomena.[6] However, the optimization of STM for high quality imaging and spectroscopy measurements is time-consuming, due to extremely sensitive tip-surface interactions, that thus limits the efficiency. This difficulty is compounded, especially in long-term measurements by physical factors such as the tip drift, piezo creep, noise, and surface states.

In the last several years, advances in machine learning (ML) have brought in significant developments to multiple areas of imaging sciences.[7-11] Recent developments in application of ML to scanning probe and electron microscopy methods have been realized to achieve autonomous experimental workflows to assist in materials discovery.[12,13] The use of neural networks and reinforcement learning based frameworks on STM systems have been demonstrated for tip functionalization and atom manipulation.[14-17]

In STM measurements, a high-quality image is necessary for an accurate determination of atomic positions, defect locations, and symmetries. An example of this is shown in **Fig. 1**, which compares STM images of graphene for different sets of control parameters. The image on the top panel (**Fig 1a**) is a higher quality image, as evidenced by the sharp peaks of the hexagonal structure in the Fourier space (**Fig. 1b**). To determine the atom-position from the scan data, we used a blob detection based on Laplacian of Gaussians (LOG) process on the noise filtered FFT-data.[9,18,19] The determined atom-position (blob), indicated by the red circles in **Fig. 1c,** shows accurate atom detection and alignment along the crystallographic axis. The lower quality image (shown in **Fig 1d**) on the other hand, introduces error in the atom position detection (shown in **Fig. 1f**). Although suitable techniques such as noise filtering and priors can be used to discern the atom-position, a bad choice of the control parameters results in noisy-images with indistinguishable features. Optimization of the instrument controls for better imaging is therefore desirable. In practice, this optimization is implemented by initially deciphering the range of values for various controls, based on prior knowledge, followed by incremental optimization, usually by trial-and-error.

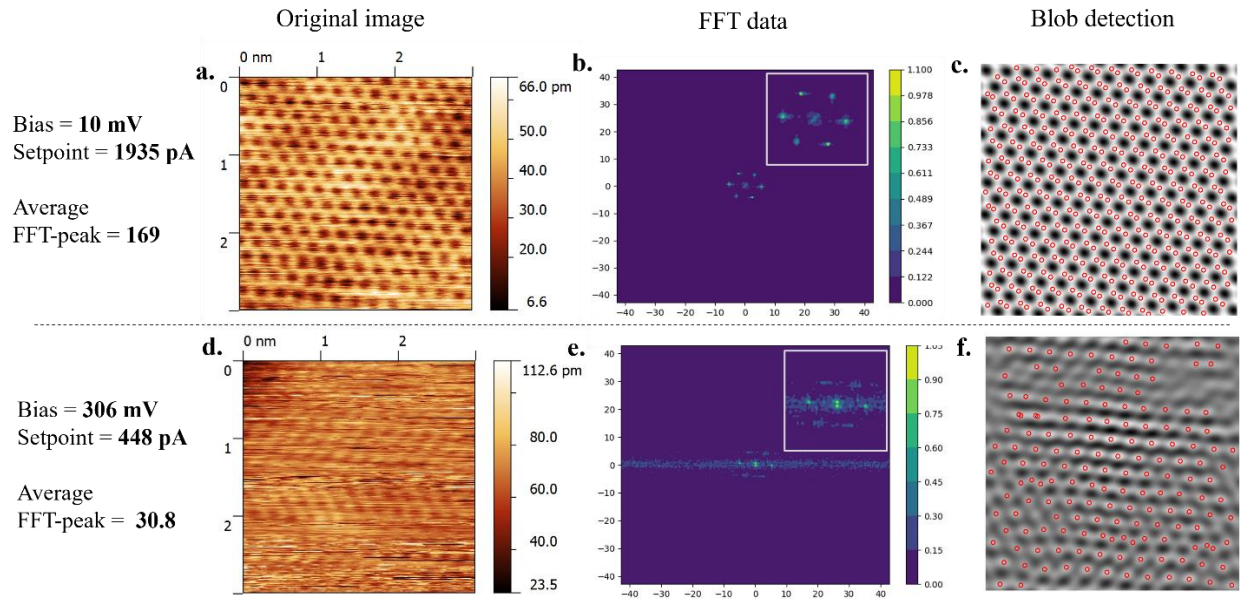

**Fig. 1**: The relevance of good quality STM images to identify atomic positions. The panel (a) in the top row corresponds to a higher quality image and this is evident in the normalized FFT data shown in (b). Inset shows the magnified central region and the six peaks are identifiable. (c) shows a blob detection method to determine the atomic positions. The detected atomic positions are indicated as circles against the grayscale background. The figure (d) in the lower panel represents lower quality STM images. This is correlated with noisier FFT image shown in (e). Inset shows the magnified central region. Using blob detection, shown in (f), introduces errors in determination of the atomic positions.

In this work we demonstrate an ML-based technique to realize autonomous convergence of the STM controls on graphene films deposited on copper substrate. The process is implemented via an active-learning based Bayesian optimization (BO) method, where a computationally cheaper Gaussian Process (GP) based regression is used to represent the target function (i.e., STM image quality) across the parameter space.[20] This process further employs an acquisition function to suggest the optimal set of STM control parameters i.e., the tip bias and the current setpoint. The suggested parameters are then directly incorporated in to the STM controls, in real time, to rapidly traverse across the parameter space. These techniques have gained traction in recent years to realize autonomous workflows[21-27].

The workflow to experimentally incorporate the BO prediction of the control parameters is illustrated in **Fig 2**. The experiment in initialed with controls that is input by the user. Once the scan is performed, the FFT data is analyzed, and the average of the six-hexagonal peak intensity (shown in **Fig. 2c**) is used as the target to be optimized.

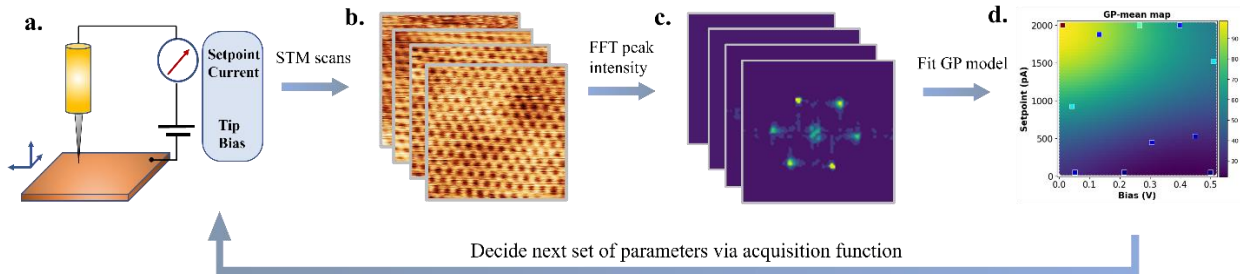

**Fig 2**: Schematic of the Automated STM control-optimization experimental workflow. (**a**) Schematic of the STM with controls being the setpoint current and the tip bias. (b) The spatial scan of graphene samples for each of the iterations are used and the (c) average of the six-FFT peaks is considered as the target for optimization. (d) A GP model is then constructed across the parameter space for using the experimental training set. Based on the GP priors, BO is used to predict the next point for subsequent experimental data sampling.

At every iteration, the preceding set of control features (i.e., current setpoint and bias) and the targets (average FFT peak intensity) serve as the training set to construct a surrogate function across the parameter space, as depicted in **Fig 2d**. The Gaussian process (GP) is utilized as the surrogate function and the output of the regression yields the predictive mean and variance ("uncertainty") values for the function of interest across the parameter space.[28] The GP surrogate function is be expressed as:

$$y = f(x) + \varepsilon$$

$$f \sim MVNormal(m, K)$$

where $\varepsilon$ is the normally distributed noise associated with the data. *MVNormal* is the multivariate normal distribution that depends on the prior mean (*m*) and the covariance matrix (*K*). The mean is usually considered to be zero and the matrix $K = k(x_i, x_j)$ depends on the covariance kernel. Once the mean and variance are mapped across the parameter space, a suitable acquisition function is employed to predict the optimal set of parameters, subsequently used for experimental-data sampling.

The acquisition function is expected to balance the process of exploration (probing where the predictive uncertainty is high) and exploitation (probing where the predicted function value is 'good'). For some acquisition functions like the upper confidence bound (UCB), the extent of exploration/exploitation strongly depends on the hand-selected hyperparameter value. We carried out experiments across a range of the hyperparameter values for the different acquisition functions.

In the experiments that were carried out, the overall ranges of tip bias and setpoint current values were set based on prior domain knowledge about the system under study and the BO was applied to find their optimal combination within those fixed ranges. Specifically, the tip-bias parameter was maintained in the range of 10-110 mV and the setpoint was in the range of 400-2000 pA. The experimental routine begins with the parameters set at 110 mV, 400 pA for the tip bias and the setpoint current respectively and continues until a set number of iterations are complete.

The experimental results indicate distinct trends associated for each of the following acquisition functions: [20,29,30]

(i) Expected Improvement (EI)

$$EI = (\mu - y_{best} - \xi)\phi\left(\frac{(\mu - y_{best} - \xi)}{\sigma}\right) + \sigma\psi\left(\frac{(\mu - y_{best} - \xi)}{\sigma}\right)$$

where, $\mu$ is the GP predictive mean, $\sigma$ is the GP predictive variance, $y_{best}$ is the maximum value of the target in the training set, $\phi$ is the cumulative distribution function and $\psi$ is the probability distribution function. $\xi$ is a slack parameter that determines the extent of improvement at every iteration.

(ii) Upper Confidence Bound (UCB)

$$UCB = \mu + \beta\sigma$$

Where $\beta$ is a factor that controls the extent of exploration/exploitation of the prediction

(iii) Probability of Improvement (PI)

$$PI = \phi\left(\frac{(\mu - y_{best} - \xi)}{\sigma}\right)$$

The dominant experimental trends for the acquisition functions is illustrated in **Fig 3**. The different rows in **Fig 3** shows the results pertaining to the acquisition function. The first column of images in **Fig 3** shows the map of the GP mean value at the end of the experiment (corresponding GP variance maps provided in **Fig S3**). The scatter points, that can be distinguished from the background, are the experimental data points. For the different acquisition functions, it can be seen the GP map indicates a higher value of the target parameter at the top left region of the parameter space which corresponds to lower values of the bias and higher values of the setpoint. This observation is in close agreement with the ground truth (**Fig S1**) dependence of the graphene samples across the given parameter space. Essentially, a lower bias and higher setpoint indicates a closer approach of the STM tip to the sample surface and results in lowering of the tunneling resistance across the tip-sample junction.

The second column in **Fig 3** corresponds to the variation of the target FFT-peak across the different iterations. The orange solid line in all the plots corresponds to the moving five-point average, which reflects the general trend of parameters sampled during the experiment. The iteration indicated by the dashed red line corresponds to a "fairly optimized" image and is denoted by a considerably higher value of the FFT parameter. This optimal condition is also correlated in the third column of figures, which indicates the variation of the bias and setpoint controls across the iterations. It is observed that the optimized FFT corresponds to low bias and high setpoint value.

For each of the set of experiments, the histogram plots (shown in **Fig S4**) were monitored, and it shows asymmetric distribution skewed towards higher setpoint and lower bias values. In the BO implementation, it is interesting to note that the prediction traverses differently across the parameter space, and the trend is unique to each of the acquisition functions and the associated hyperparameter value.

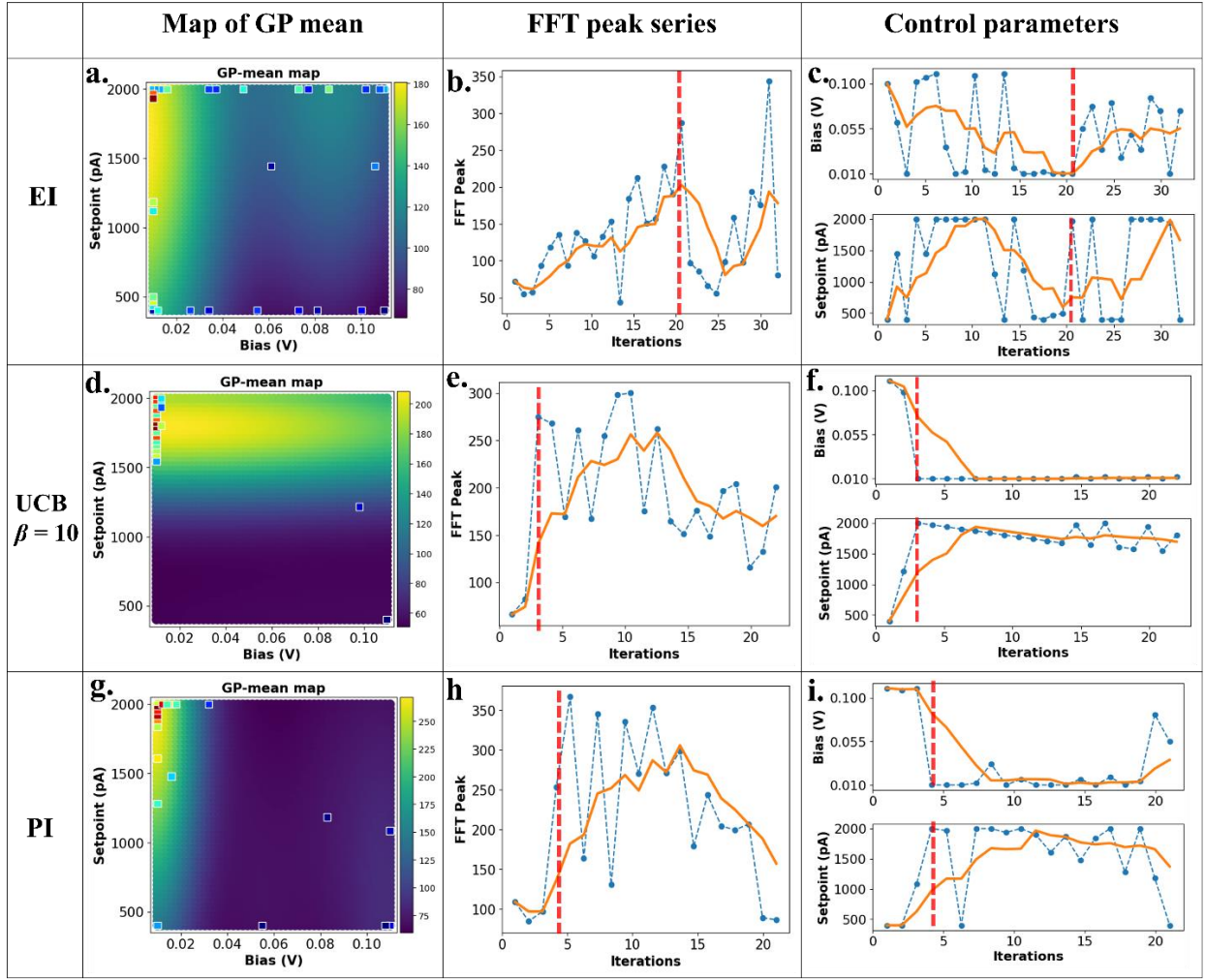

**Fig. 3**: Summary of the trends observed for experiments with different acquisition functions, namely Expected Improvement, EI: (a – c). Upper confidence bound, UCB: (d-f), and Probability of Improvement. PI: (g-i). The first column represents the GP predictive-mean map, and the scattered points are the experimental data. The second column contains the variation of the FFT-peak parameter across the iterations. The orange solid line in the plots is the five-point moving average of the scatter points. The dashed red vertical line represents the iteration where prediction corresponds to the optimized scan quality. The third column contains the variation of the control parameters: bias and setpoint across the different iterations.

The results summarizing the trends for the acquisition functions is tabulated in **Table 1**. In the case of the EI and the PI function, optimization of the parameters is efficient, and this evident in the histogram plots (**Fig. S4d, S4l**). However, the prediction continues to explore points away from the optimal region.

| Acquisition Function | Hyperparameter | Optimized at iteration: | Inference |
|---|---|---|---|
| **Expected Improvement (EI)** | $\xi = 0$ | 3, 8 | Explorative |
| | $\xi = 0.001$ | 4, 20 | |
| | $\xi = 0.1$ | 4, 8 | |
| **Upper Confidence Bound (UCB)** | $\beta = 1 - 5$ | Exploited at false maxima | Dominantly Exploitative |
| | $\beta = 10 - 100$ | < 10 iterations, exploits around (false) maxima | Explorative at high $\beta$ |
| | $\beta = 500 - 100$ | < 10 iterations, continues exploration | |
| **Probability of Improvement (PI)** | $\xi = 0, 0.001, 0.1$ | < 10 iterations | Explores in the neighborhood of the maxima. |

**Table 1**: Summary of the results for experiments for the different acquisition functions. For each of the acquisition functions, experimental trends were studied over a range of hyperparameter values.

In contrast, the UCB function is dominantly exploitative for low and intermediate values of $\beta$. The predictions are in close vicinity of the optimal value. We observed that low values of $\beta = 1 - 5$ leads to the predictions around a local (or even false) maxima, and in some cases does not converge even for iterations ~ 100 (**Fig S5**). However, for higher values of $\beta$ (in the range 500 - 1000), the prediction is explorative, with no sign of convergence (**Fig S6**). For the EI and the PI acquisition functions, the trends do not differ dramatically for the different $\xi$ values (**Fig S7**, **S8**).

We utilize the strong dependence of the UCB function on $\beta$ (and in conjunction with other acquisition functions) to develop two policies for autonomous convergence of STM scan controls.

1. Combination of Acquisition functions:

Here we have used the EI and UCB functions as the two agents for exploration and exploitation respectively. Given that EI ($\xi = 0.1$) predicts optimized control parameter within the first few iterations (indicated in **Table 1**), the policy here uses the initial ten iterations for exploration with the EI followed by the next ten iterations of UCB ($\beta = 1$) for convergence.

The results of the experiments are summarized in the **Fig 4**. **Fig 4a** shows the GP predictive mean (variance shown in **Fig S9b**) which shows a higher density of prediction in the optimal region of the parameter space. In **Fig 4b** and **4c**, the blue shaded region indicates the part of the experiment with EI acquisition function, followed by the white background denoting the utilization of UCB function. The EI is explorative with higher variations in the FFT values (**Fig 4b**) and the predicted parameters (**Fig 4c**). However, once the UCB agent is deployed, the FFT values steadily converges to the optimal value. This is correlated with the convergence of the control parameters to high setpoint and low bias values, as is evident in **Fig 4c** and the histogram plots in **Fig 4d**.

A similar experiment was carried out with the combination of the PI and UCB acquisition functions, and the convergence show similar trend (**Fig S10**).

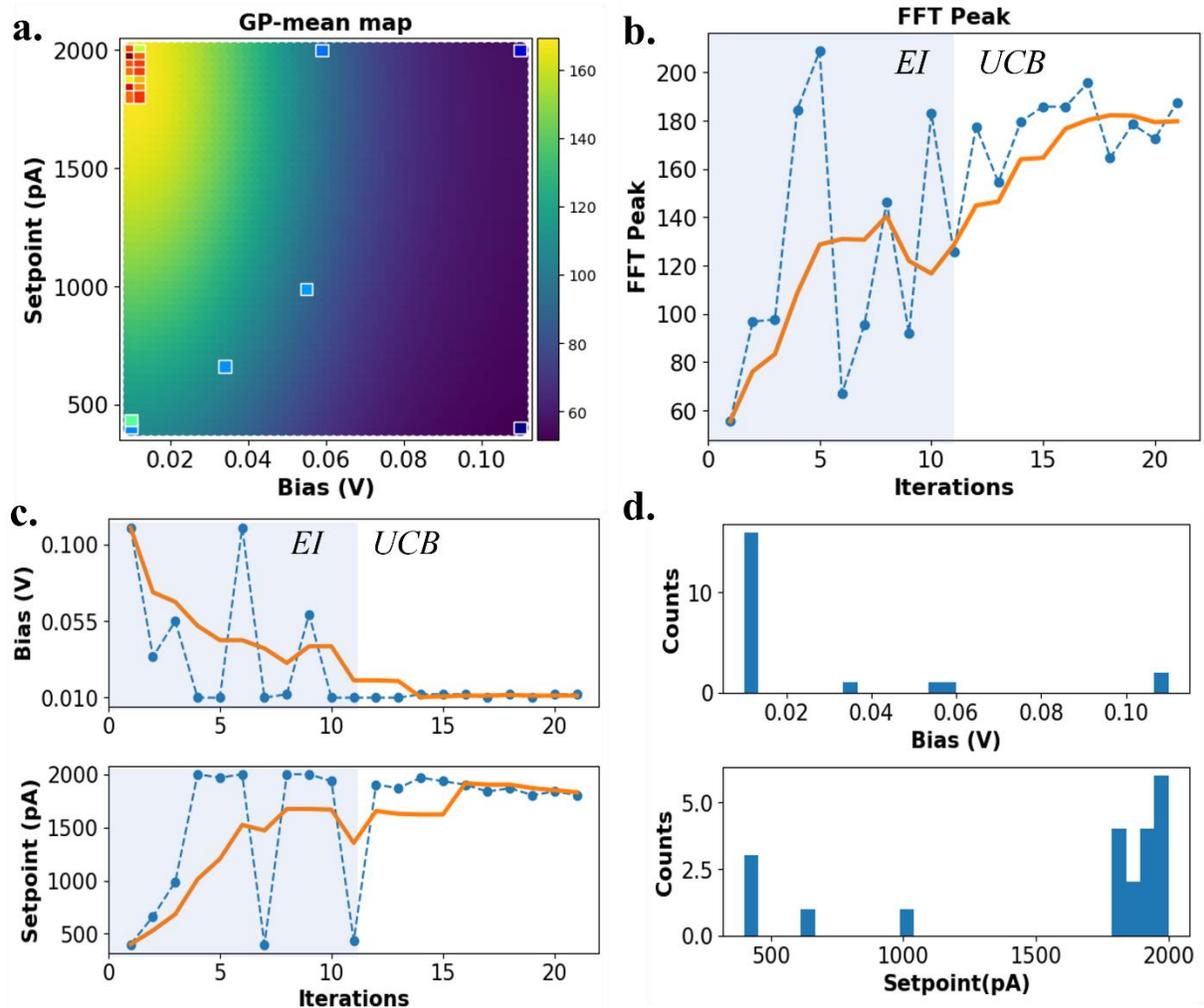

**Fig. 4**: Plots describing autonomous convergence using the combination of EI and UCB acquisition functions. (a) shows the GP mean map across the parameter space. (b) shows the variation of the variation of the FFT peak parameter across the iterations. The orange solid line is the five-point moving average of the scatter data points. The blue shaded region corresponds to the EI ($\xi = 0.1$) acquisition function and the white region corresponds to the UCB ($\beta = 1$). (c) shows the variation of the control parameters across the iterations. (d) shows the histogram plots of the acquisition function at the end of the iterations.

2. $\beta$-annealing.

In the second policy, we employ the UCB acquisition function while gradually reducing the $\beta$ parameter to transition from exploration to exploitation. In this experiment, we extend the control-parameter space across the working region of the STM, i.e., bias in the range of 10–510 mV and setpoint in the range of 50–2000 pA. The experiment starts with an initial $\beta = 1000$ for dominant exploration. As the experiment proceeds, the $\beta$ value is "annealed" by iteratively reducing the value by 10 %, in comparison to the previous iteration. We therefore expect the prediction to be "greedy" in the later part of the experiment. The results of the experiments are illustrated in **Fig 5**.

**Fig 5a** shows the map of the GP mean and the scatter points show that predictions spanned across the parameter space (variance map shown in **Fig S9d**). The top panel in **Fig 5b** shows that variation of the $\beta$ parameter across the iterations starting from $\beta = 1000$ in the initial iteration to $\beta \sim 5$ in the final iteration. The variation of the target FFT parameter (orange solid line) shows a gradual improvement in the initial part of the experiment, followed by a saturation in the later part of the iteration. This is correlated with the control parameter values (in **Fig 5c**) which show high variability followed by convergence to higher setpoint and lower bias values. This prediction trends are correlated with the asymmetry in the histogram plots shown in **Fig 5d**.

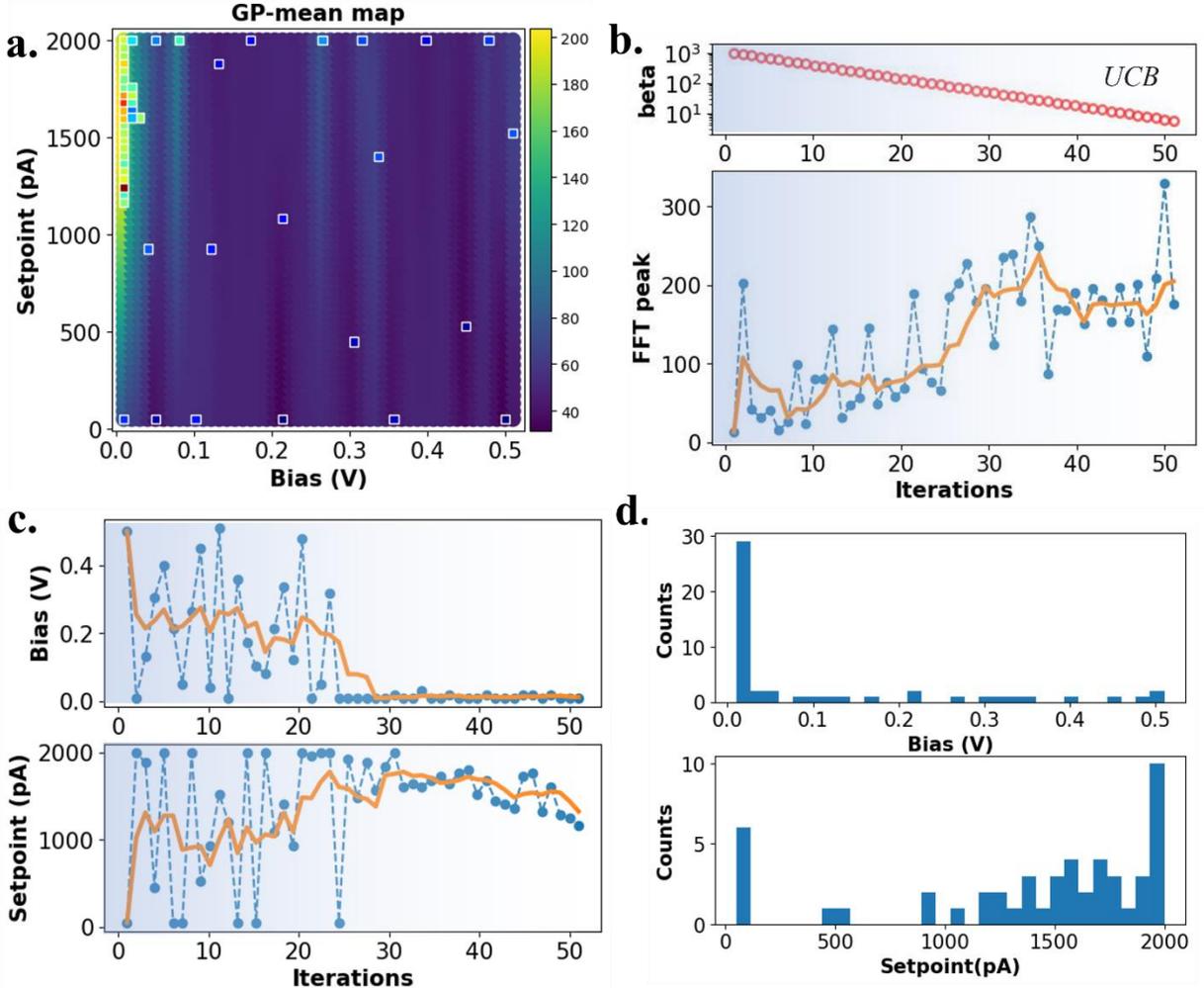

**Fig. 5**: Plots describing autonomous convergence using the β-annealing technique associated with the UCB acquisition functions. (a) shows the GP mean map across the parameter space. The top panel in (b) shows the β-profile that is steadily reduced during the experiment. The lower plot in (b) shows of the variation of the FFT peak parameter across the iterations. The orange solid line in the plots is the five-point moving average of the scatter points. (c) shows the variation of the control parameters across the iterations. (d) shows the histogram plots of the acquisition function at the end of the experiment.

This above-described policies are demonstrations of autonomous convergence of scan parameters. In this experiment we maintained a relatively slow *β-annealing* to allow for exploration across the larger parameter space. This rate of exploration/exploitation can however be tuned by suitably modifying the range and the annealing rate of the *β* value.

In the multiple experiments that were carried out, the image quality is sensitive to the tip stability. The dynamic nature of tip-surface interactions due to processes such as temperature fluctuations, piezo creep, and the presence of atomic impurities influences the image quality, despite fixing the control parameters. To account for this, we estimated a "FFT-variance-parameter" which was found to have to be inversely correlated with the tip stability. The details of the estimation, analysis and impact of tip conditioning is described in Section 9 of the Supplementary Information. Further we also note that given the sensitive nature of the tip, the correlation between image stability and quality would not hold true under conditions that results in modification of the tip. This could include unintentional functionalization or the presence of impurities at the tip.

In conclusion, we demonstrate the autonomous convergence of the STM scan controls in real-time which is implemented by integrating the instrumentation feedback, on-the-fly data analysis, and BO-based prediction of the control parameters. This work demonstrates the proof-of-concept on the graphene sample, which has a relatively smoother dependence across the energy landscape. The BO procedure can be extended dynamically to complex systems which show deviation from the standard convex trends, where regions of high image quality and stability can exist as isolated regions within the parameter space.

The workflow methodology presented in this work can be modified and applicable to systems such as semiconductors, 2D layers, topological materials, and other quantum structures, as well as other imaging techniques. Depending on the system under investigation, suitable physical models can be incorporated into the GP regression. The convergence policies when further combined with spectroscopic techniques, allows for autonomous exploitation/exploration of target properties such as superconducting phases, topological defects, impurities, and interfacial regions.

## Methods

**Sample Preparation and STM imaging**

STM experiments were conducted on CVD-grown graphene monolayers on copper foil. Samples were cleaned by sonication in an acetone bath for 20 minutes and annealed in ultra-high vacuum ($<1 \times 10^{-10}$ Torr) at 500 C for 4 hours to remove the contaminants on the graphene surface. After annealing, samples were transferred to in-situ STM system (Omicron VT-STM). All measurements were performed in constant current mode using chemically etched Tungsten tips. Quality and metallicity of tips were confirmed on Cu(100) surface before the measurements. The sample temperature was maintained at 302.5 K during the experiment to reduce thermal fluctuations. In our experiments, the tip is grounded, and a negative of the indicated bias was applied to the sample. This preserves the convention of the tip-bias indicated in this work. For each STM image, a 3 nm × 3 nm area is scanned with a scan speed of 20 nm/s.

**Code and Data availability**

The data and the code for the Bayesian Optimization program implemented on the STM is provided at: https://github.com/gnganesh99/BO-for-AutoSTM

**Data Analysis**

*Bayesian Optimization:*

The autonomous workflow of the STM experiments to find the optimal imaging condition has been done with Bayesian optimization (BO), which is designed to provide a derivative free optimization task of an expensive (time/resources) black-box problem, where the functional representation between the input (control parameters) and the output (experimental results) is unknown and the optimal solution can be found with minimal experiments. This is implemented at each iteration where the control features (i.e., current setpoint and bias) and the targets (average FFT peak intensity) are used to a surrogate function across the parameter space. Here a Gaussian process (GP) is utilized as the surrogate function and the output of the regression yields the predictive mean and variance ("uncertainty") values for the function of interest across the parameter space. The GP surrogate function can be expressed as:

$$y = f(x) + \varepsilon$$

$$f \sim MVNormal(m, K)$$

where $\varepsilon$ is the normally distributed noise associated with the data. *MVNormal* is the multivariate normal distribution that depends on the prior mean ($m$) and the covariance matrix ($K$). The mean is usually considered to be zero and the matrix $K = k(x_i, x_j)$ depends on the covariance kernel. In this study we used the radial basis function (RBF) kernel since we expect smoothness of the associated image quality (and therefore the FFT-peak) across the parameter space. The form of RBF kernel is given as:

$$k_{RBF}(x_i, x_j) = \eta \exp\left(\frac{-d(x_i, x_j)^2}{2l^2}\right)$$

Where $d$ is the Euclidean distance, $\eta$ is the output scale and $l$ is the kernel length scale. Given a set of input parameters $X_*$, the posterior mean ($\mu^{post}$) and variance ($\sigma^{post}$) is given by:

$$f_* \sim MVNormal(\mu_\theta^{post}, \sigma_\theta^{post})$$

$$\mu_\theta^{post} = m(X_*) + K(X_*, X|\theta)K(X, X|\theta)^{-1}(y - m(X))$$

$$\sigma_\theta^{post} = K(X_*, X_*|\theta) - K(X_*, X|\theta)K(X, X|\theta)^{-1}K(X, X_*|\theta)$$

where $\theta = (\eta, l)$ are the inferred kernel parameters. Once the mean and variance are mapped across the parameter space, a suitable acquisition function is employed to predict the optimal set of parameters. This point of acquisition-function-maxima is considered for subsequent experimental-data sampling. The experiment is automated until a set number of iterations are complete. The optimization process is driven by the objective to maximize the STM image quality, which is indicated by the average FFT peak intensity.

The Bayesian Optimization framework have been developed in "Pytorch" Python environment. The Gaussian process regression is modelled with "Gpytorch" library package. The hyper-parameter optimization for GP training have been conducted with Adam optimizer with learning rate 0.1, and computational epochs = 150 at every iteration.

*Noise Filtering*:

Prior to the blob detection, the FFT images were filtered for noise using low pass and a high pass filters. A threshold of 0.6 for the top panel (and 0.3 for the bottom panel of Fig. 1) was implemented on the normalized FFT to pick the dominant peaks in the FFT data. An inverse FFT operation was performed to reconstruct the atomic structure.

*Blob detection*:

The blob detection was carried out based on the Laplacian of Gaussian (LOG) function, that is available on the scikit-image library.

*Peak identification:*

Peaks in the normalized FFT data were identified using the find_peaks function available in the scipy,signal library. The following function descriptions were used to identify the maximum number of peaks: *height* = 0.05, *width* = 1, *threshold* = 0.02. The six peaks were identified by carrying on rotation operations on the images and discerning the quadrant and the lattice parameter associated with the peak positions.

**Author Contributions:**

GN designed the LabVIEW controls, performed the experiments, and compiled the data. SH arranged for the sample and setup the instrumentation for the STM experiments. AB worked on the Bayesian Optimization code used in the experiments. RV and MZ conceptualized, coordinated, and supervised the implementation of the autonomous STM. All authors were involved in the data analysis and contributed towards manuscript preparation.

**Acknowledgments**

The work was supported by the Center for Nanophase Materials Sciences (CNMS), which is a US Department of Energy, Office of Science User Facility at Oak Ridge National Laboratory. We thank Prof. Sergei V. Kalinin (University of Tennessee) for helpful discussions about optimizing imaging parameters in scanning probe and scanning tunneling microscopy experiments. We also acknowledge Mr. Sai Mani Prudhvi Valleti for helping with the blob-detection method.

# Supplementary Information

# Autonomous convergence of STM control parameters using Bayesian Optimization

Ganesh Narasimha[1], Saban Hus[1], Arpan Biswas[1], Rama Vasudevan[1]*, Maxim Ziatdinov[2]*

[1] *Center for Nanophase Material Sciences (CNMS), Oak Ridge National Laboratory (ORNL), Oak Ridge, Tennessee, USA – 37831*

[2] *Computational Sciences and Engineering Division (CSED), Oak Ridge National Laboratory (ORNL), Oak Ridge, Tennessee, USA – 37831*

## 1. Ground truth of average FFT-peak intensity.

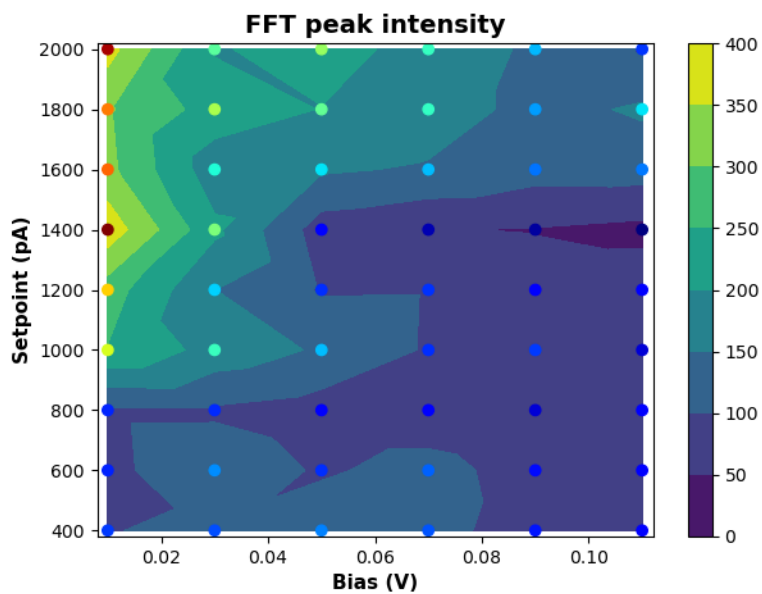

**Fig S1**: The ground truth variation of the average FFT-peak intensity across the parameter space. Here the tip bias varies in the range of 10 – 110 mV and the setpoint is in the range 400 – 2000 pA. The scatter points represent the data where at experiments were carried out.

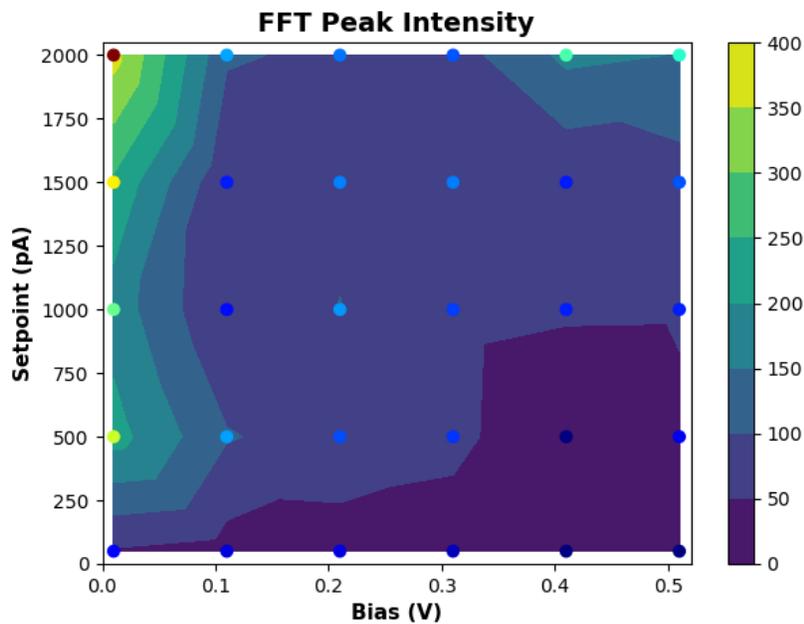

**Fig S2**: The ground truth variation of the average FFT-peak intensity across the larger parameter space. Here the tip bias varies in the range of 10 – 510 mV and the setpoint is in the range 50 – 2000 pA. The scatter points represent the data where at experiments were carried out.

## 2. GP predictive mean and corresponding variance maps

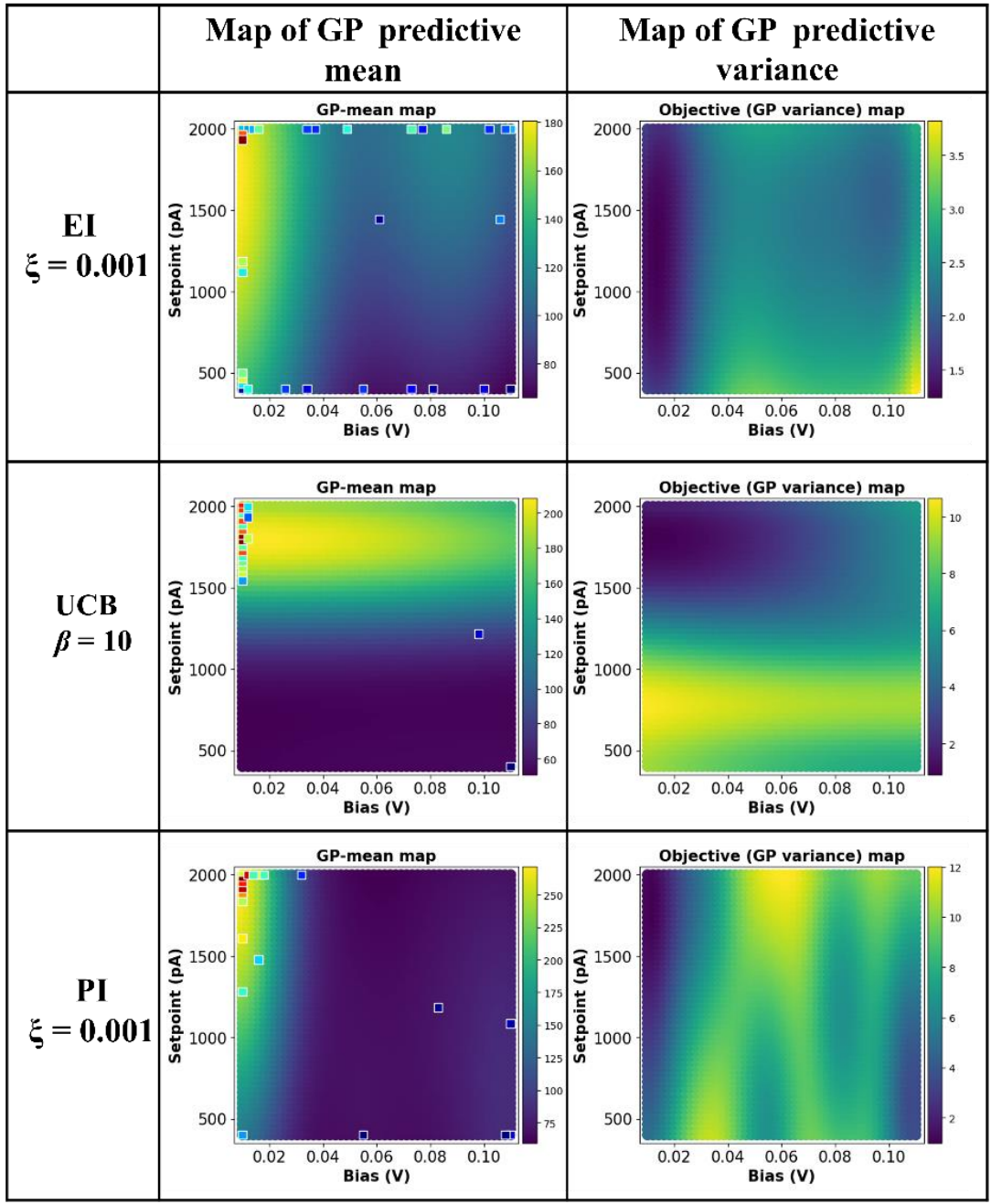

**Fig S3**: First column: The map of the GP predictive mean illustrated in Fig 3 of the text for the different acquisition functions. The second column shows the corresponding GP predicted variance maps across the parameter space.

## 3. Variation of the data exploration in the parameter space with the histogram plots

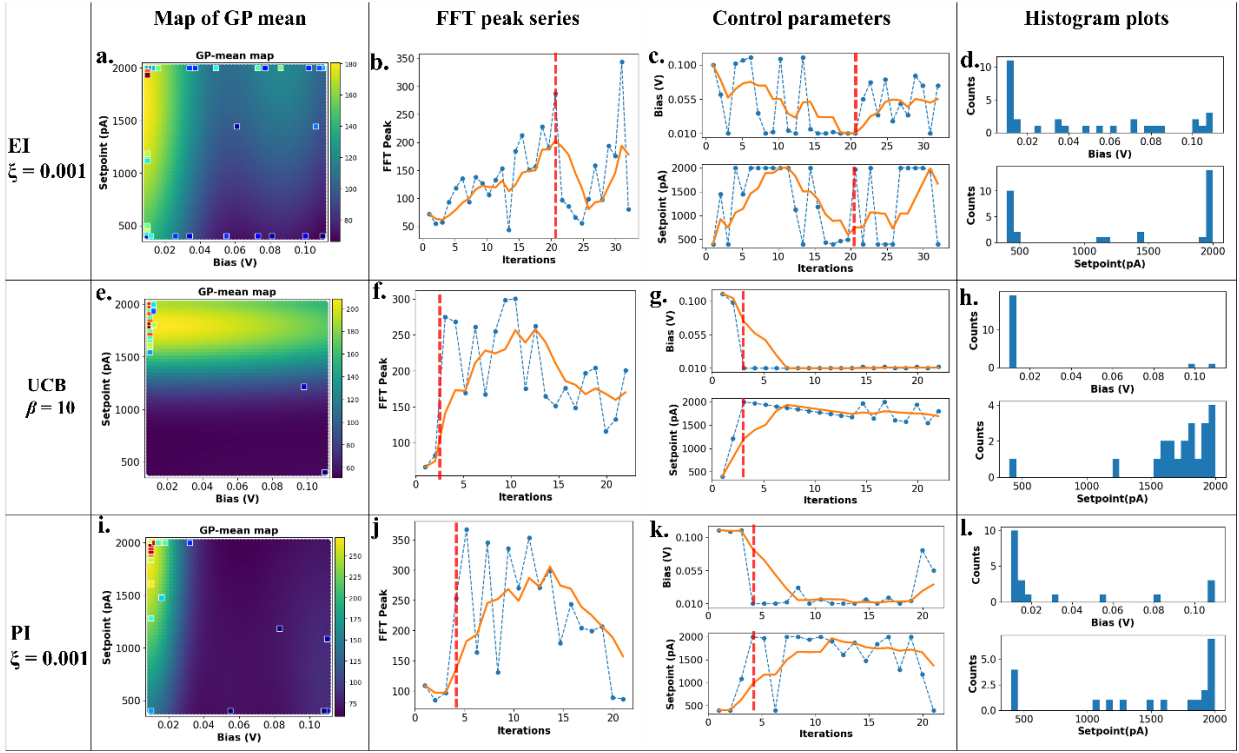

**Fig. S4**: Summary of the trends observed for experiments with different acquisition functions, namely Expected Improvement, EI: **(a – d)**. Upper confidence bound, UCB: **(e-h)**, and Probability of Improvement. PI: **(i-l)**. The first column represents the GP predictive-mean map, and the scattered points are the experimental data. The second column contains the variation of the FFT-peak parameter across the iterations. The orange solid line in the plots is the five-point moving average of the scatter points. The dashed red vertical line represents the iteration where prediction corresponds to the optimized scan quality. The third column contains the variation of the control parameters: bias and setpoint across the different iterations. The fourth column represents the histogram plots of the control parameters at the end of the experiment.

## 4. Automated experiments using UCB with variation of *β*-parameter.

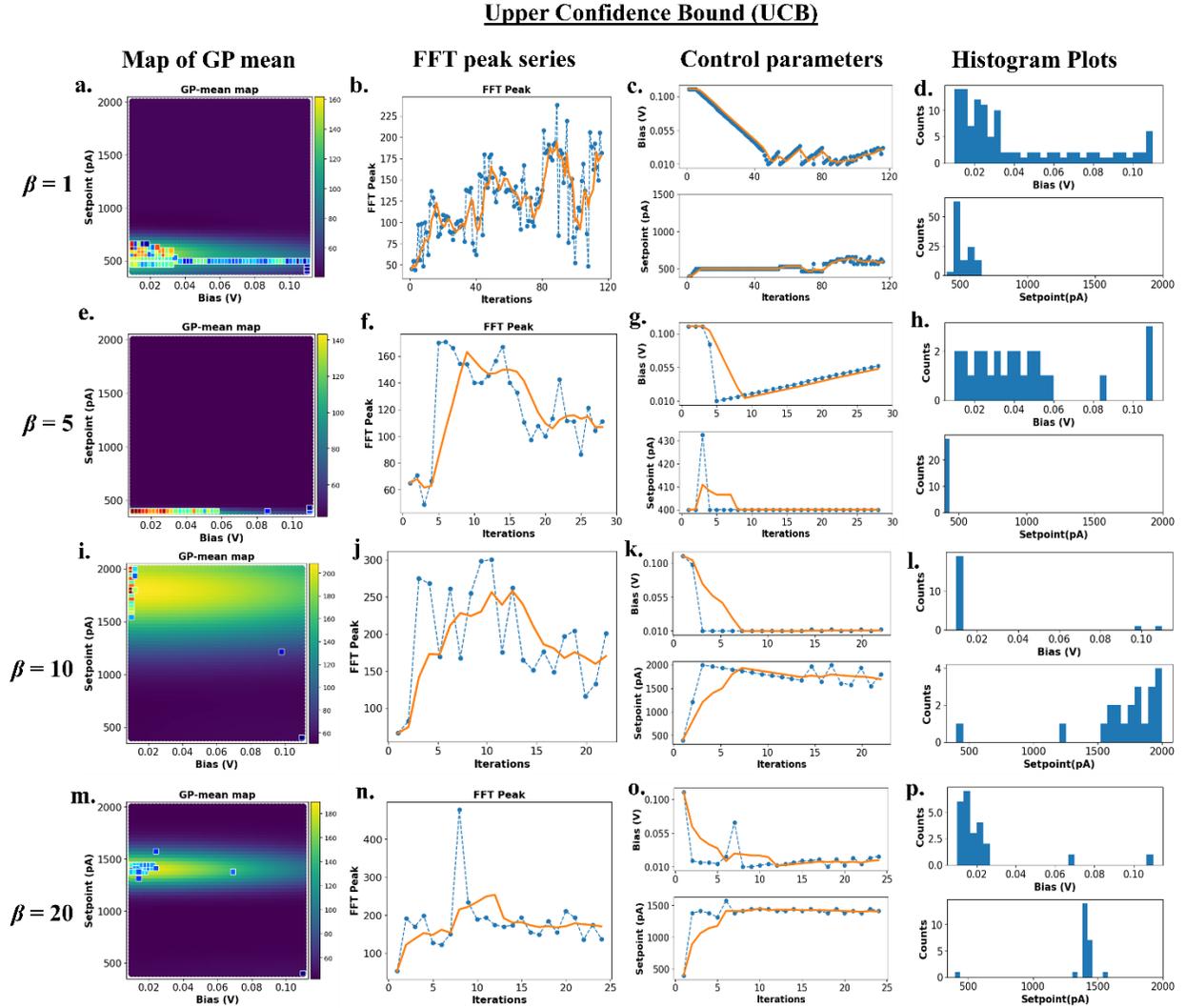

**Fig. S5**: Summary of the trends observed for experiments with UCB as the acquisition functions, with **(a – d)** *β* = 1, **(e-h)** *β* = 5, **(i-l)** *β* = 10, and **(m-p)** *β* = 20. The first column represents the GP predictive-mean map, and the scattered points are the experimental data. The second column contains the variation of the FFT-peak parameter across the iterations. The orange solid line in the plots is the five-point moving average of the scatter points. The third column contains the variation of the control parameters: bias and setpoint across the different iterations. The fourth column represents the histogram plots of the control parameters at the end of the experiment.

**Upper Confidence Bound (UCB)**

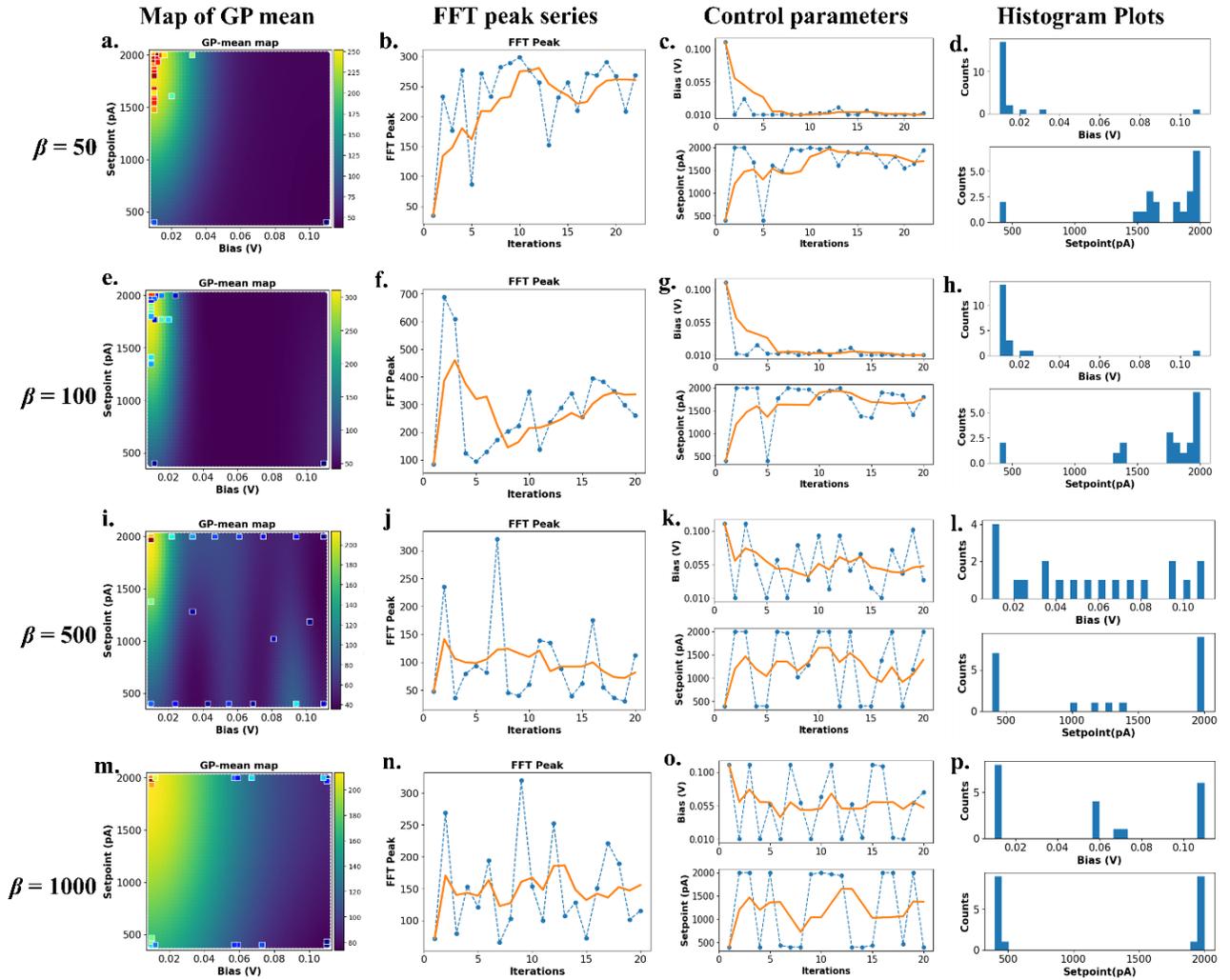

**Fig. S6**: Summary of the trends observed for experiments with UCB as the acquisition functions, with **(a – d)** $\beta = 50$, **(e-h)** $\beta = 100$, **(i-l)** $\beta = 500$, and **(m-p)** $\beta = 1000$. The first column represents the GP predictive-mean map, and the scattered points are the experimental data. The second column contains the variation of the FFT-peak parameter across the iterations. The orange solid line in the plots is the five-point moving average of the scatter points. The third column contains the variation of the control parameters: bias and setpoint across the different iterations. The fourth column represents the histogram plots of the control parameters at the end of the experiment.

## 5. Automated experiments using EI with variation of ξ-parameter.

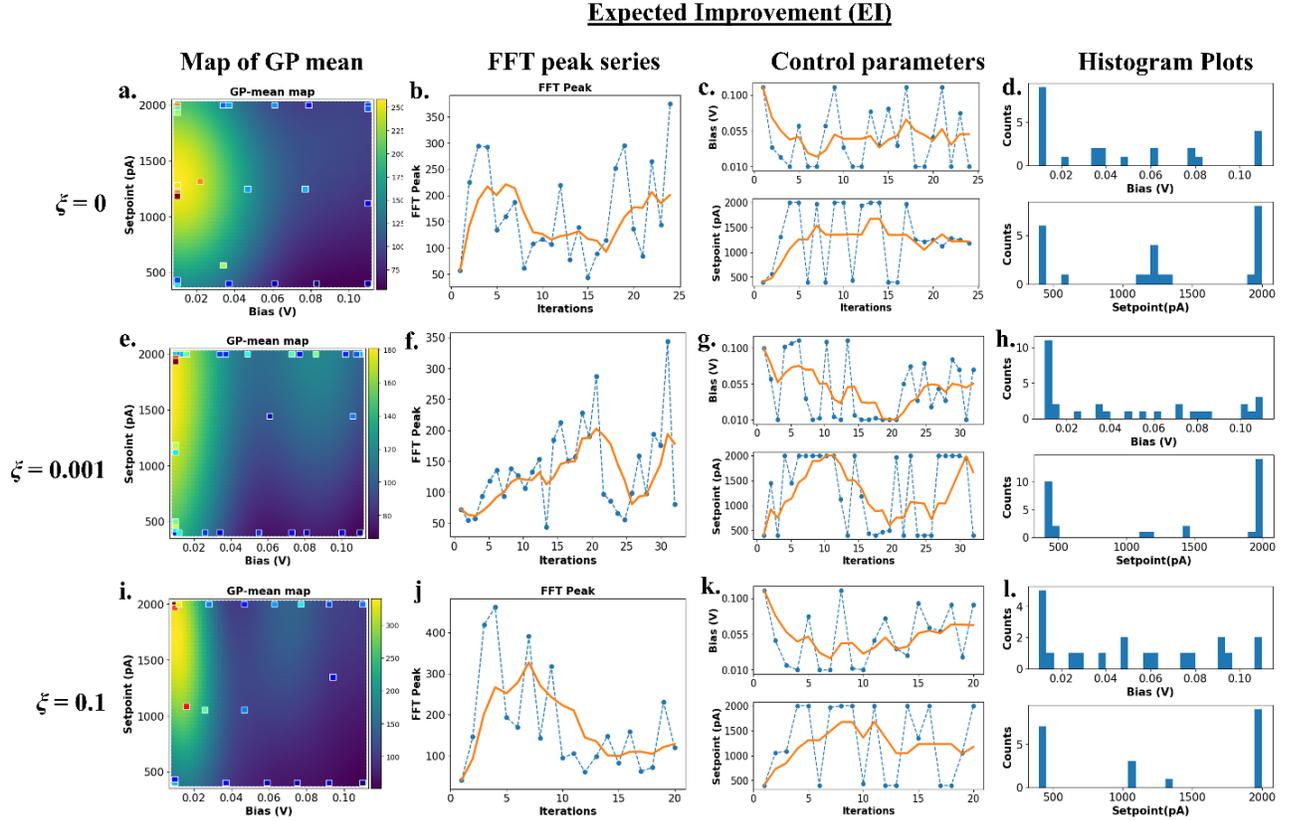

**Fig. S7**: Summary of the trends observed for experiments with EI as the acquisition functions, with **(a – d)** ξ = 0, **(e-h)** ξ = 0.001, **(i-l)** ξ = 0.1. The first column represents the GP predictive-mean map, and the scattered points are the experimental data. The second column contains the variation of the FFT-peak parameter across the iterations. The orange solid line in the plots is the five-point moving average of the scatter points. The third column contains the variation of the control parameters: bias and setpoint across the different iterations. The fourth column represents the histogram plots of the control parameters at the end of the experiment.

## 6. Automated experiments using PI with variation of $\xi$-parameter.

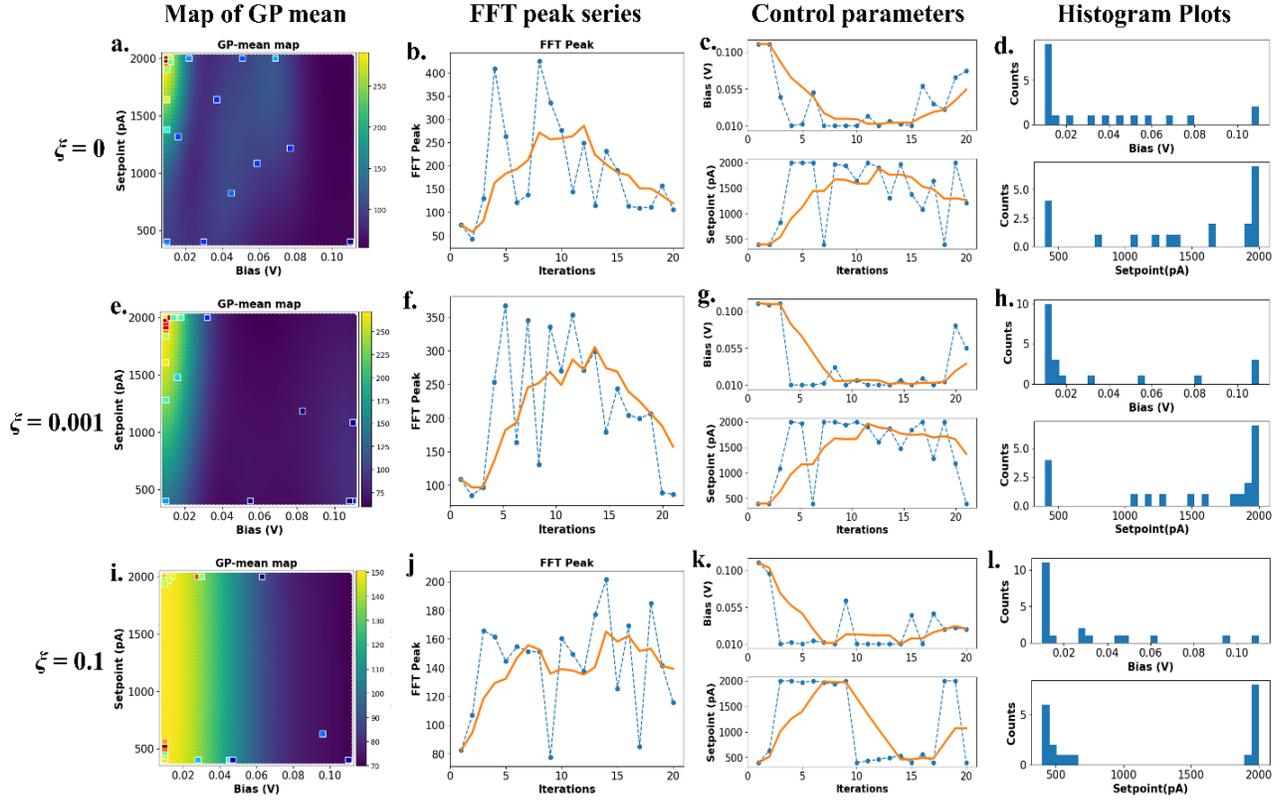

**Fig. S8**: Summary of the trends observed for experiments with PI as the acquisition functions, with **(a – d)** $\xi = 0$, **(e-h)** $\xi = 0.001$, **(i-l)** $\xi = 0.1$. The first column represents the GP predictive-mean map, and the scattered points are the experimental data. The second column contains the variation of the FFT-peak parameter across the iterations. The orange solid line in the plots is the five-point moving average of the scatter points. The third column contains the variation of the control parameters: bias and setpoint across the different iterations. The fourth column represents the histogram plots of the control parameters at the end of the experiment.

## 7. GP mean and variance maps for convergence policies.

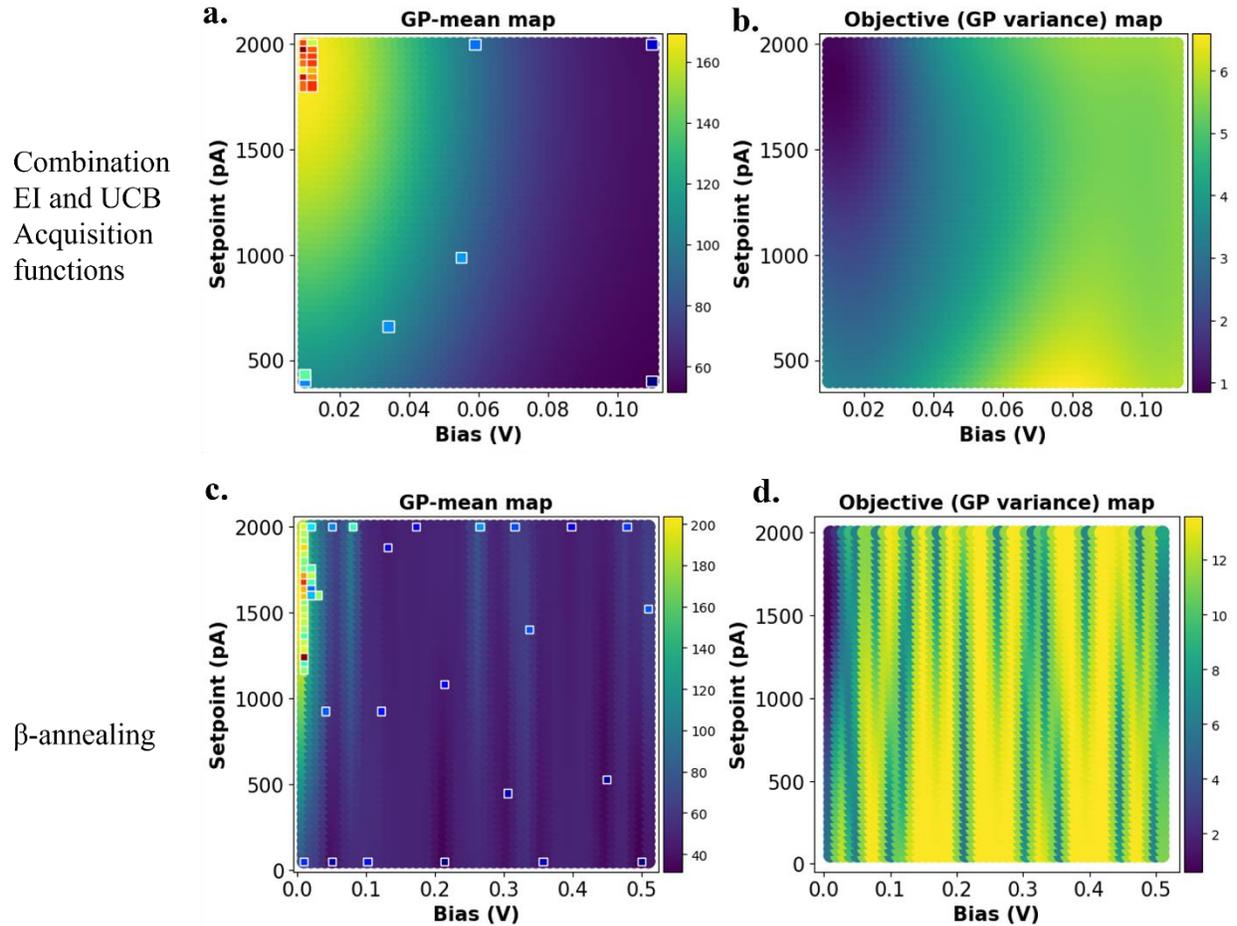

**Fig S9**: GP mean and variance maps for the convergence policies: (**a, b**): Combination of the acquisition functions EI and UCB; and (**c,d**): β-annealing policy. The description of the policies is provided in Fig 4 and Fig 5 of the main text

## 8. Policy for autonomous convergence: Combination of PI and UCB

This experiment involves a convergence policy where a combination of acquisiton functions: PI ($\xi = 0.1$) and UCB ($\beta = 5$) were employed. The PI was deployed in the initial seven iterations followed by the UCB in the next seven iterations.

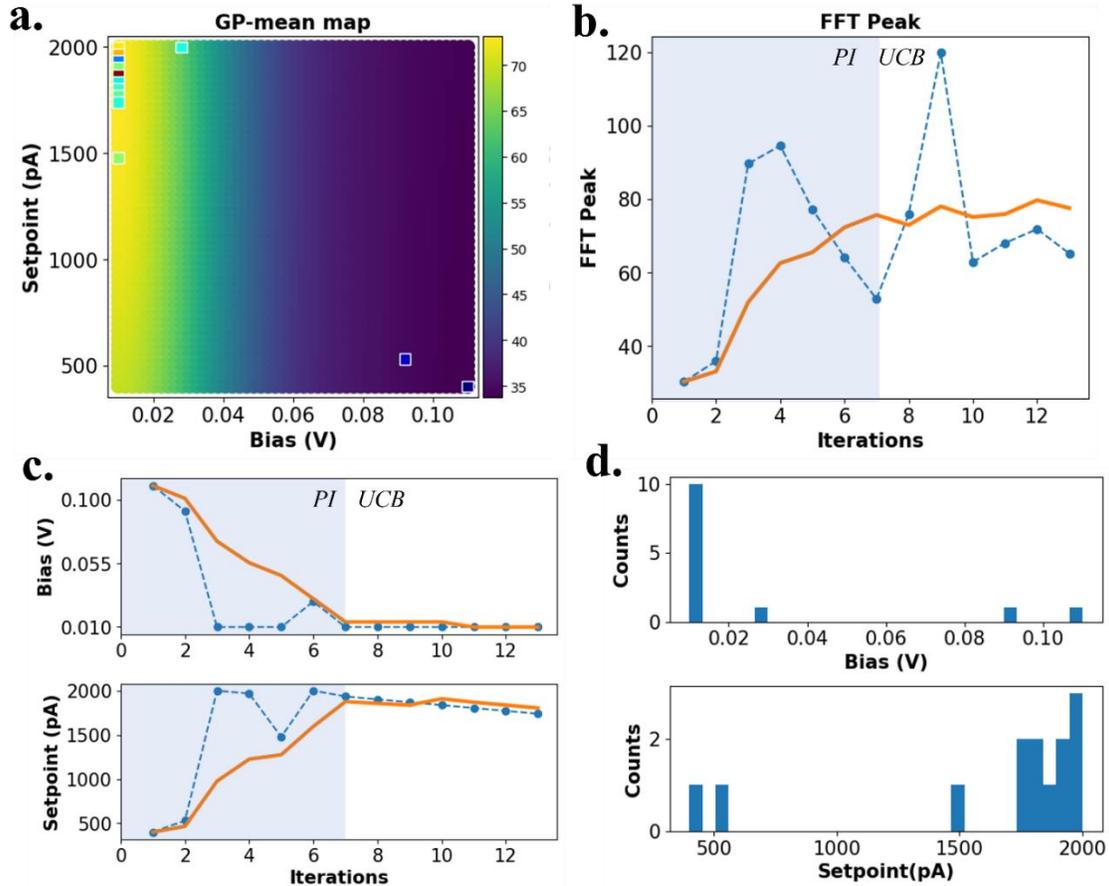

**Fig. S10**: Plots describing autonomous convergence using the combination of PI and UCB acquisition functions. (a) shows the GP mean map across the parameter space. (b) shows the variation of the variation of the FFT peak parameter across the iterations. The orange solid line is the five-point moving average of the scatter data points. The blue shaded region corresponds to the PI ($\xi = 0.1$) acquisition function and the white region corresponds to the UCB ($\beta = 5$). (c) shows the variation of the control parameters across the iterations. (d) shows the histogram plots of the acquisition function at the end of the iterations.

## 9. Tip Stability.

One of the crucial concerns related to the STM measurements and optimization is in relation to the tip stability. The STM measurements are influenced by the extremely sensitive tip-surface interactions. The dynamic nature of tip-surface interactions due to processes such as temperature fluctuations, piezo creep, and the presence of atomic impurities can influence the image quality, despite fixing the control parameters.

In addition of the quantification of the image quality, we estimate a "FFT-variance-parameter" that indicates the extent of the tip stability. The schematic for this estimation is explained in **Fig S11a**. Essentially, the scan image is fragmented into (time) series segments and the prominent FFT peak is analyzed for each of the segments. The variance in the FFT-peak parameter is correlated with the tip stability inherent during the scanning process.

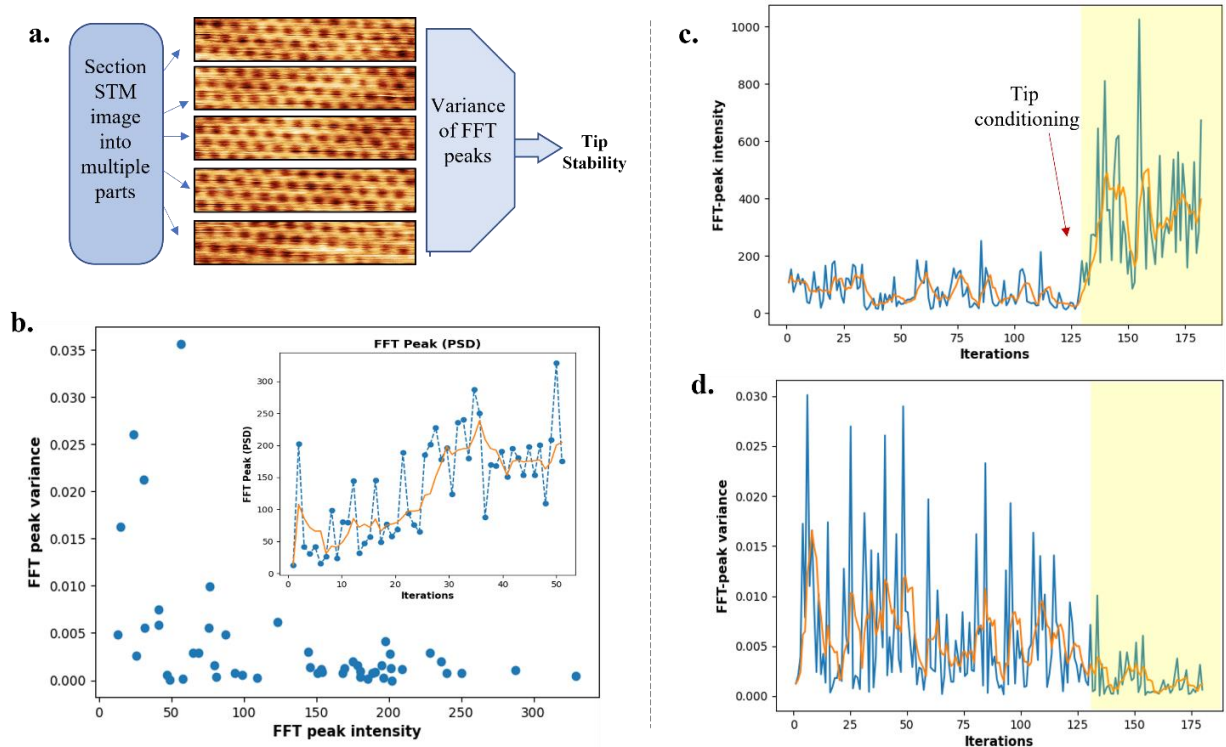

**Fig. S11**: Tip stability and correlation. (a) shows the methodology to determine the FFT-peak variance to estimate scan stability. A scan data is fragmented to multiple sections and the prominent FFT peak is determined for each of them. The variance across the different sections is used to estimate tip stability. (b) Correlation between FFT-peak variance of a scan data and the FFT-peak intensity. Inset shows the series experiment corresponding to the β-annealing experiment. The orange solid line in the plots is the five-point moving average of the scatter points. (c) and (d) shows variation of FFT-peak value and FFT-peak-variance, respectively, over 180

scans with random sampling across the parameter space. The second highlighted part which corresponds to the STM scans after tip-cleaning procedure denotes higher values for the FFT-peak parameter. This corresponds to the reduction of the FFT-peak-variance shown in (d).

A higher stability in the scanning condition corresponds to a lower variance in the FFT-peak values. Further, we find that the variance parameter has an inverse correlation with the FFT-peak value, as shown in **Fig S11b**, which implies improved stability under optimized scan conditions. Given the sensitive nature of the tip, the correlation between image stability and quality would not hold true under conditions that results in modification of the tip. This could include unintentional functionalization or the presence of impurities at the tip.

The tip stability can be improved by additional techniques such as tip cleaning and conditioning. For the "tip conditioning", a tip-biasing pulse of 5 V was applied over a period of 30 ms. This was followed by a large area scan (200 nm). The desired frame width was reduced gradually over multiple scans.

**Fig S11c** and **d** shows the data for FFT peak and peak-variance, respectively, for experiments where the control parameters are sampled randomly across the parameter space. Prior to beginning the second part of the experiment (indicated as the highlighted region), the tip is "conditioned" by a bias-pulsing procedure followed by a large area scan. It is observed that the mean FFT-peak value increases (**Fig. S11c**), and this correlated to a reduction of the peak-variance parameter shown in **Fig S11d**. This trend establishes the correlation between image quality and scan stability.

In effect, the BO based approach for improved image quality indirectly optimizes the scan stability. However, implementation of tip-conditioning procedures is advisable to improve the imaging conditions.